\documentclass[twocolumn,trackchanges, times]{aastex62}
\hypersetup{linkcolor=red,citecolor=green,filecolor=cyan,urlcolor=magenta}

\newcommand{\rqq}{\textquotedblright}
\newcommand{\lqq}{\textquotedblleft}

\graphicspath{{./}{figures/}}

\submitjournal{\textit{Accepted for publication in ApJ}}

\shorttitle{Ly$\alpha$ Intensity Mapping at $z=5.7$ and $6.6$}

\shortauthors{Kakuma et al.}
\begin{document}
\title{SILVERRUSH. IX. Ly$\alpha$ Intensity Mapping with Star-Forming Galaxies at $z=5.7$ and $6.6$:
A Possible Detection of Extended Ly$\alpha$ Emission 
at $\gtrsim$100 comoving kpc around and beyond the Virial-Radius Scale of Galaxy Dark Matter Halos
}

\author{Ryota Kakuma}
\affil{Institute for Cosmic Ray Research, The University of Tokyo, 5-1-5 Kashiwanoha, Kashiwa, Chiba 277-8582, Japan}

\author{Masami Ouchi}
\affiliation{National Astronomical Observatory of Japan, 2-21-1 Osawa, Mitaka, Tokyo 181-8588, Japan}
\affiliation{Institute for Cosmic Ray Research, The University of Tokyo, 5-1-5 Kashiwanoha, Kashiwa, Chiba 277-8582, Japan}
\affiliation{Kavli Institute for the Physics and Mathematics of the Universe (Kavli IPMU, WPI), The University of Tokyo, 5-1-5 Kashiwanoha, Kashiwa, Chiba 277-8583, Japan}

\author{Yuichi Harikane}
\affiliation{National Astronomical Observatory of Japan, 2-21-1 Osawa, Mitaka, Tokyo 181-8588, Japan}

\author{Yoshiaki Ono}
\affiliation{Institute for Cosmic Ray Research, The University of Tokyo, 5-1-5 Kashiwanoha, Kashiwa, Chiba 277-8582, Japan}

\author{Akio K. Inoue}
\affiliation{Department of Physics, School of Advanced Science and Engineering, Faculty of Science and Engineering, Waseda University, 3-4-1 Okubo, Shinjuku, Tokyo 169-8555, Japan}
\affiliation{Waseda Research Institute for Science and Engineering, Faculty of Science and Engineering, Waseda University, 3-4-1 Okubo, Shinjuku, Tokyo 169-8555, Japan}

\author{Yutaka Komiyama}
\affiliation{National Astronomical Observatory of Japan, 2-21-1 Osawa, Mitaka, Tokyo 181-8588, Japan}
\affiliation{Graduate University for Advanced Studies (SOKENDAI), 2-21-1 Osawa, Mitaka, Tokyo 181-8588, Japan}

\author{Haruka Kusakabe}
\affiliation{Observatoire de Gen\`{e}ve, Universit\'{e} de Gen\`{e}ve, 51 Ch. des Maillettes, 1290 Versoix, Switzerland}

\author{Chien-Hsiu Lee}
\affiliation{NSF's National Optical-Infrared Astronomy Research Laboratory, 950 North Cherry Avenue, Tucson, AZ 85742, USA}

\author{Yuichi Matsuda}
\affiliation{National Astronomical Observatory of Japan, 2-21-1 Osawa, Mitaka, Tokyo 181-8588, Japan}
\affiliation{Graduate University for Advanced Studies (SOKENDAI), 2-21-1 Osawa, Mitaka, Tokyo 181-8588, Japan}

\author{Yoshiki Matsuoka}
\affiliation{Research Center for Space and Cosmic Evolution, Ehime University, 2-5 Bunkyo-cho, Matsuyama, Ehime 790-8577, Japan}

\author{Ken Mawatari}
\affiliation{Institute for Cosmic Ray Research, The University of Tokyo, 5-1-5 Kashiwanoha, Kashiwa, Chiba 277-8582, Japan}

\author{Rieko Momose}
\affiliation{Department of Astronomy, Graduate School of Science, The University of Tokyo, 7-3-1 Hongo, Bunkyo, Tokyo 113-0033, Japan}

\author{Takatoshi Shibuya}
\affiliation{Kitami Institute of Technology, 165 Koen-cho, Kitami, Hokkaido 090-8507, Japan}

\author{Yoshiaki Taniguchi}
\affiliation{The Open University of Japan, Wakaba 2-11, Mihama-ku, Chiba 261-8586, Japan}

\begin{abstract}
We present results of the cross-correlation Ly$\alpha$ intensity mapping 
with Subaru/Hyper Suprime-Cam (HSC) ultra-deep narrowband images 
and Ly$\alpha$ emitters (LAEs) at $z=5.7$ and $6.6$ in a total area of $4$ deg$^2$. 
Although overwhelming amount of data quality controls have been performed for the narrowband images, 
we further conduct extensive analysis evaluating systematics of 
large-scale point-spread-function wings, sky subtractions, and unknown errors 
based on physically uncorrelated signals and sources found in real HSC images and object catalogs, respectively. 
Removing the systematics, we carefully calculate cross-correlations between Ly$\alpha$ intensity 
of the narrowband images and the LAEs. 
We tentatively identify very diffuse Ly$\alpha$ emission with the $\simeq 3\sigma$ ($\simeq 2\sigma$) significance 
at $\gtrsim$ 100 comoving kpc (ckpc) far from the LAEs at $z=5.7$ ($6.6$), 
around and probably even beyond a virial radius of star-forming galaxies with $M_\mathrm{h}\sim10^{11}M_\odot$. 
The diffuse Ly$\alpha$ emission possibly extends up to $1$,$000$ ckpc 
with the surface brightness of $10^{-20}$--$10^{-19}$ erg s$^{-1}$ cm$^{-2}$ arcsec$^{-2}$. 
We confirm that the small-scale ($<150$ ckpc) Ly$\alpha$ radial profiles of LAEs 
are consistent with those obtained by recent MUSE observations. 
Comparisons with numerical simulations suggest that 
the large-scale ($\sim150$--$1$,$000$ ckpc) Ly$\alpha$ emission are not explained 
by unresolved faint neighboring galaxies including satellites, 
but by a combination of Ly$\alpha$ photons emitted from the central LAE and other unknown sources, 
such as cold-gas streams and galactic outflow. 
We find no evolution in the Ly$\alpha$ radial profiles of our LAEs from $z=5.7$ to $6.6$, 
where theoretical models predict a flattening of the profile slope made by cosmic reionization, 
albeit with our moderately large observational errors.
\end{abstract}


\section{Introduction} \label{sec:intro}
Ly$\alpha$ emission is one of the strongest emission lines in galaxy spectra.
The Ly$\alpha$ emission is thus widely used for various studies in astronomy, such for galaxy formation, the circumgalactic medium (CGM), the intergalactic medium (IGM), and large scale structures.

Ly$\alpha$ emitters (LAEs) have been identified up to $z$ $\sim$ 9 by deep imaging surveys (e.g. \citealp{Kashikawa2006, Ouchi2008, Ouchi2010, Hu2010a, Konno2014a, Matthee2015a, Santos2016a, Zheng2017} and spectroscopic observations (e.g. \citealp{Deharveng2008, Adams2011a, Cassata2015a, Zitrin2015a, Song2016, Stark2017}).
Statistical studies have revealed the general picture of LAEs' physical properties.
Typical LAEs represent low-mass high-$z$ galaxies \citep{Ono2010, Ono2010a, Kashikawa2012, Harikane2018a}, some of which are thought to be candidates of population III galaxies \citep{Partridge1967,Schaerer2003}.

Ly$\alpha$ photons are produced in the recombination processes of ionized hydrogen ({\sc Hii}) gas, and resonantly scattered in neutral hydrogen ({\sc Hi}) gas surrounding a galaxy \citep{Smith2018}.
In the last two decades, extended Ly$\alpha$ emission around galaxies has been identified (e.g., \citealt{Hayashino2004,Matsuda2012}).
\citet{Steidel2011} have found that high-$z$ star-forming galaxies on average show extended Ly$\alpha$ haloes (LAHs) around them based on stacking analyses of narrowband (NB) imaging data. 
\citet{Momose2014a} have made samples of 100--3600 LAEs at $z = 2.2$--$6.6$ from NB imaging data, and conducted image stacking with intensive tests for checking the systematics.
The LAHs at $z=2.2$--$6.6$ have been identified with exponential scale lengths of $\sim$ $5$--$10$ physical kpc (pkpc).
The Multi-Unit Spectroscopic Explorer (MUSE) on the Very Large Telescope enables us to detect LAHs on the individual basis with no stacking analysis (\citealt{Wisotzki2016}; \citealt{Leclercq2017a}; \citealt{Leclercq2020}).
\citet{Leclercq2017a} report the detection of LAHs around 145 individual LAEs at $3 \leq z \leq 6$ in the Hubble Ultra Deep Field (HUDF).
\citet{Leclercq2017a} have conducted two exponential component decomposition of a core and a halo for the Ly$\alpha$ radial profile on pseudo NB images of the MUSE data, and found 80\% of objects show the radial profile of the Ly$\alpha$ more extended than the one of the UV continuum.
Extended Ly$\alpha$ emission have been found not only around  LAEs but also around other types of objects.
\citet{Martin2014} and \citet{Cantalupo2014} have revealed quasars with Ly$\alpha$ emission extended to $250$--$460$ pkpc.
This largely-extended Ly$\alpha$ emission is attributed to fluorescent emission from central quasars.

Recently, studies of intensity mapping analyses have investigated the large-scale matter distribution by measuring the integrated emission from galaxies and the IGM. (\citealt{Kovetz2017}; see also \citealt{Chang2010}; \citealt{Croft2016}; \citealt{Croft2018a}; \citealt{Chiang2019}).
\citet{Croft2016, Croft2018a} derive a cross-correlation function between the positions of quasars and Ly$\alpha$ intensity in Sloan Digital Sky Survey spectra of luminous red galaxies after subtracting best-fitting model galaxy spectra, and detect a signal around quasars on scales of $1$--$15\ h^{-1}$ comoving Mpc.
Quasars are very rare objects unlike galaxies. The strong radiation of quasars ionize the IGM gas around quasars. The environment around quasar is special in the universe. For understanding the galaxy formation, it is important to explore the environment around galaxies.
To detect the diffuse emission from large-scale matter distribution around galaxies, we can make use of data from wide-field surveys and the intensity mapping analysis. 
In this study, we exploit wide and deep optical Hyper Suprime-Cam (HSC; \citealp{Miyazaki2012}; see also \citealp{Miyazaki2018, Komiyama2018,  Furusawa2018, Kawanomoto2018}) images obtained by the Subaru Strategic Program (HSC SSP; \citealp{Aihara2018a}). 
The HSC SSP survey expends 300 nights of Subaru observing time over 5 years since 2014. The survey consists of three layers; Wide (W), Deep (D), and UltraDeep (UD). In the D and UD layers, NB imaging is carried out with 4 filters (\textit{NB387}, \textit{NB816}, \textit{NB921}, \textit{NB1010}).
These NB imaging data allow us to make a large LAE sample and to use as Ly$\alpha$ 2D intensity maps.

In this paper, we will report a first marginal detection of the Ly$\alpha$ emission around star-forming galaxies largely extended around and probably even beyond the dark matter halo (DMH) virial radius ($r_\mathrm{vir} \sim 150$ comoving kpc (ckpc) for $M_\mathrm{h} \sim 10^{11} M_\odot$) at $z =5.7$ and $6.6$ based on the intensity mapping analysis of the cross correlation between Ly$\alpha$ intensity maps and the positions of LAEs.
Using the large Ly$\alpha$ 2D intensity maps and the large LAE samples obtained by the  HSC-SSP NB survey, we investigate the large-scale matter distribution around general star-forming galaxies.
This paper has the following structure.
In Section 2, we describe our LAE catalogs and images. 
Section 3 presents our analysis and results. 
In Section 4, we conduct a test to evaluate systematic errors. 
We briefly discuss a physical origin of extended Ly$\alpha$ emission and cosmic reionization in section 5.
We summarize our findings in Section 6.
Throughout this paper we use AB magnitudes \citep{Oke1983} and adopt the Planck cosmological parameter sets of the TT, TE, EE+lowP+lensing+ext result \citep{PlanckCollaboration2016}: 
$\Omega_\mathrm{m}=0.3089$, 
$\Omega_\mathrm{\Lambda}=0.6911$,
$\Omega_\mathrm{b} = 0.049$,
$h = 0.6774$.
In this cosmology, $1 ''$ corresponds to transverse sizes of 40 (42) ckpc at $z = 5.7$ ($6.6$).

\section{Data} \label{sec:data}
\begin{figure}[htbp!]
\plotone{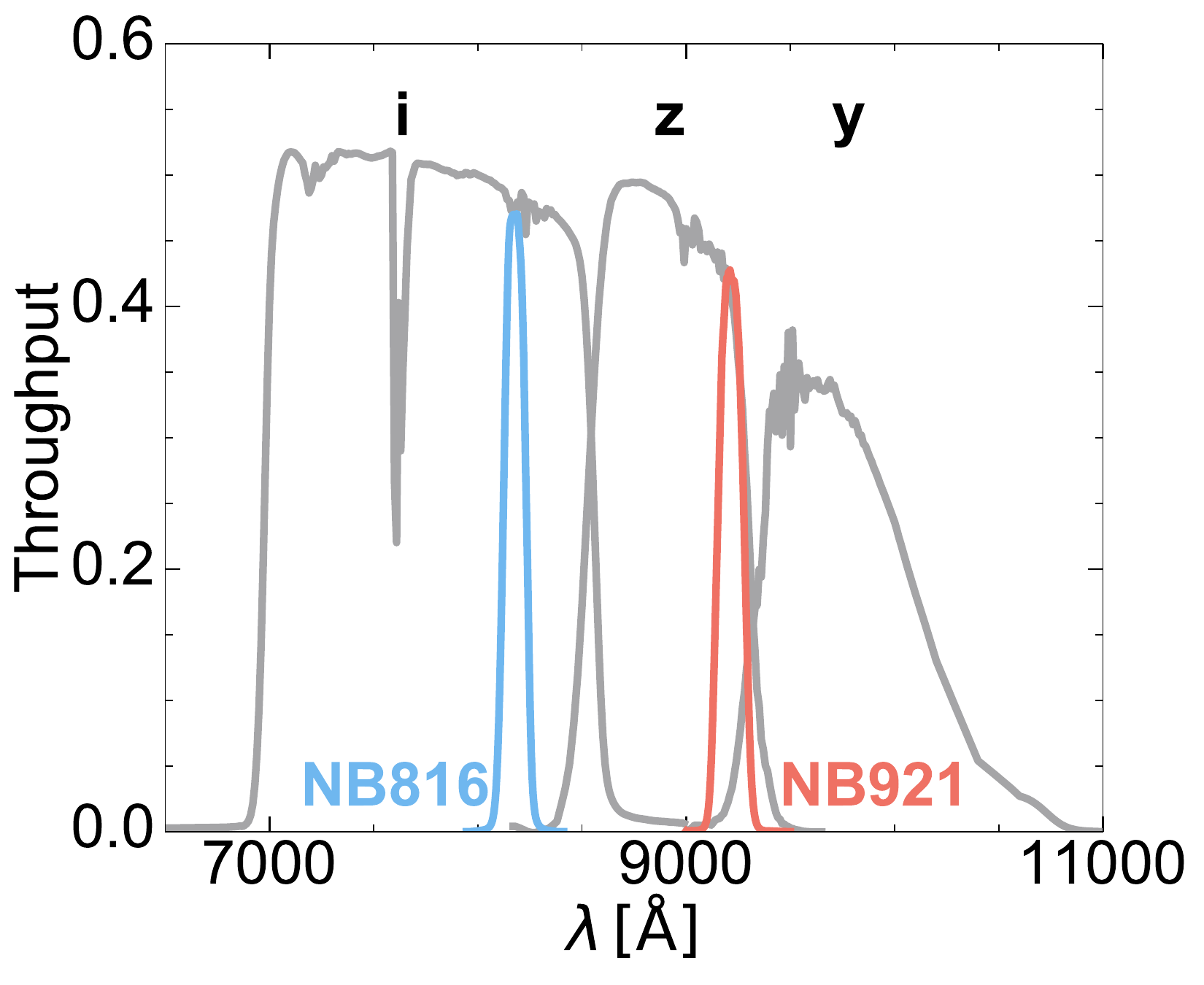}
\caption{
HSC filter transmission curves of the narrowband and broadband filters. 
The blue (red) line represents the transmission curve of the \textit{NB816} (\textit{NB921}) filter. 
The gray lines denote the HSC $i$, $z$, and $y$ broadband filters.
}
\label{fig:filtercureve}
\end{figure}

In this study, we exploit the HSC-SSP NB survey data that is taken with \textit{NB816} and \textit{NB921} filters \citep{Ouchi2018}.
The \textit{NB816} (\textit{NB921}) filter has a central wavelength of $\lambda_{c} = 8177$\AA\ ($9215$\AA) and an FWHM of $\Delta\lambda=113$\AA\ ($135$\AA).
The \textit{NB816} and \textit{NB921} filters can identify redshifted Ly$\alpha$ emission at $z = 5.726 \pm 0.046$ and $6.580 \pm 0.056$, respectively.
The HSC narrowband filter transmission curves are shown in Figure \ref{fig:filtercureve}.
The depth of NB image data is $\sim 25.5$ mag and the FWHM size of the point spread function in the HSC images is typically $\sim 0\farcs8$ (see more details in \citealp{Shibuya2018a}).

\subsection{LAE Catalogs}
For source catalogs of our intensity mapping analysis, we use the LAE catalogs obtained in the program of ``Systematic Identification of LAEs for Visible Exploration and Reionization Research Using Subaru HSC'' (SILVERRUSH; \citealp{Ouchi2018}, \citealp{Shibuya2018a}). 
The SILVERRUSH LAE catalogs is derived from the HSC-SSP survey data whose first data release is presented in \citet{Aihara2018b}.
This is the largest $z \gtrsim 5$ LAE catalog to date.
The details of the SILVERRUSH catalogs are listed in Table \ref{table:SILVERRUSH}.

\begin{figure*}[htbp!]
\plotone{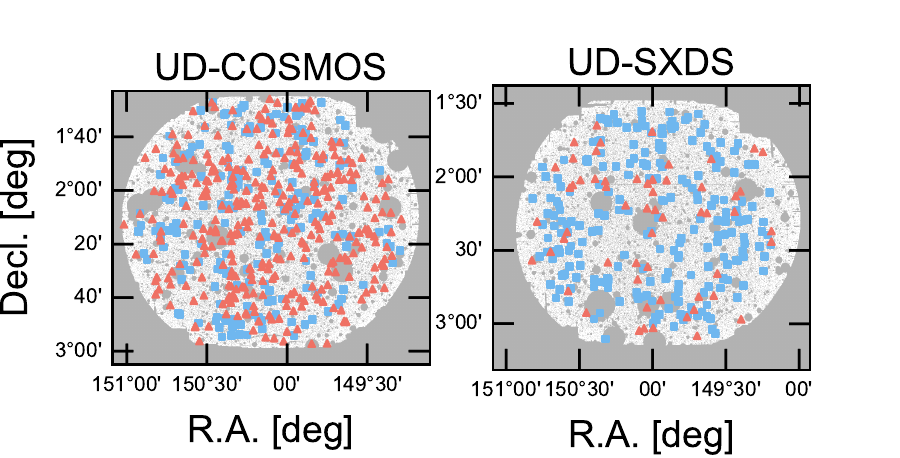}
\caption{
Sky distribution of LAEs in the UD-COSMOS and UD-SXDS fields taken from the SILVERRUSH catalogs.
The blue squares and red triangles represent positions of LAEs at $z=5.7$ and $6.6$, respectively.
}
\label{fig:UD_Fields}
\end{figure*}

The number of $z=6.6$ LAEs in UD-SXDS field appears to be smaller than that of the UD-COSMOS field. \citet{Shibuya2018a} attribute this difference to the seeing size of the \textit{NB921} images of the UD-SXDS field ($\sim 0\farcs7$) that is worse than the one of the UD-COSMOS field ($\sim 0\farcs6$). 
Figure 5 of \citet{Shibuya2018a} shows that the surface number density of LAEs in the UD-COSMOS field is comparable to those identified in Subaru/Suprime-Cam NB surveys.
Though the SILVERRUSH program exploits the HSC-SSP data taken in the D and UD layers, we use only those taken in the UD layer that provides the highest quality data in the HSC SSP survey.
Figure \ref{fig:UD_Fields} displays the sky distribution of the LAEs in the UD layer.

\begin{table}[htbp]
  \centering
   \begin{tabular}{ccccc} \hline
     Field & Area & $m_{\rm lim}$ & $N_{\rm LAE}$ & $\log\left(L_{\mathrm{Ly\alpha}}/\mathrm{\,[erg\,s^{-1}]}\right)$  \\ 
    & ($\mathrm{deg}^2$)  & (mag) & & \\ 
    (1) & (2) & (3) & (4) & (5) \\ \hline \hline
    \multicolumn{5}{c}{\textit{NB816} ($z$ $\sim 5.7$) }\\ \hline
     UD-COSMOS & 1.97  & 25.7  &  201  & - \\ 
     UD-SXDS & 1.93 & 25.5 &  224 & -  \\ 
     Total & 3.9 & - & 425  & 42.0 - 43.8 \\ \hline
     \multicolumn{5}{c}{\textit{NB921} ($z$ $\sim 6.6$) }\\ \hline
     UD-COSMOS  & 2.05 & 25.6  &  338  & -  \\ 
     UD-SXDS & 2.02 & 25.5  &   58 &-  \\ 
     Total & 4.07 & - & 396  & 42.3 - 44.0\\ \hline
   \end{tabular} 
  \caption{
  (1) Field. 
  (2) Survey Area. 
  (3) Limiting magnitude of the NB image defined by the 5$\sigma$ detection level in a $1\farcs5$ diameter circular aperture. 
  (4) Number of the LAEs. 
  (5) The range of LAE Ly$\alpha$ luminosity.}
\label{table:SILVERRUSH}
\end{table}

\subsection{Images}
\label{subsec:Images}
For an intensity map of the Ly$\alpha$ emission at the redshift same as those of LAEs at $z = 5.7$ ($6.6$), we use \textit{NB816} (\textit{NB921}) images in the HSC-SSP survey S18A data release \citep{Aihara2019}.
The NB images in the S18A data release are $\sim 1 $ mag deeper than those in the S16A data release.
The images in the S18A data release are reduced with the HSC pipeline v6.7 (\citealp{Bosch2018}) that implements an improved sky background subtraction approach.
The HSC pipeline v6.7 jointly models and subtracts scattered lights from bright objects and instrumental features crossing all CCDs.
In addition, the sky background emission are estimated and subtracted in the mosaic with a grid size of 6000 pixel, corresponding to about 17 arcmin. 
This new sky subtraction method reduces overfittings and oversubtractions of the sky background made by small scale fluctuations.
Although the latest S18A data is released very recently (June 2018), we have not selected LAEs from this data.

The NB images contain not only Ly$\alpha$ emission at $z = 5.7$ or $6.6$ but also continuum and emission lines that are emitted from low-$z$ sources.
The effects from these contaminants can nevertheless be removed by taking a spatial cross-correlation with LAEs because these sources randomly located on the sky with respect to the LAEs.
The low-$z$ sources only add noise on the cross-correlation between the LAEs and Ly$\alpha$ emission.

Note that we need to mask bright sources on the images to improve the signal to noise ratio (S/N) of the diffuse extended emission.
The HSC pipeline sets some flags to each pixel.
We do not use pixels with flags of BRIGHT\_OBJECT or DETECTED. 
The BRIGHT\_OBJECT flag is given to pixels where nearby very bright objects would affect to the background subtraction or detection.
The DETECTED flag is given to the pixels in which sources are detected at the S/N$>5$ level.
With these flags, the pixels in which LAEs are detected are also masked out.
We focus on extended Ly$\alpha$ emission at $\geq 1\farcs5$ separated from the LAEs.
This is about two times larger than the size of LAEs on the images that are marginally resolved.

\section{Analysis \& Results} \label{sec:analysis}
\subsection{Cross-correlation analysis}
\label{subsec:cross-correlation}
We derive the angular cross-correlation function between the LAEs and the Ly$\alpha$ intensity $\xi_{\rm IM}(r)$, taking a mean over all LAE-pixel pairs that are separated by $r$ within a certain bin:
\begin{equation}
\xi_{\mathrm{IM}}(r) = \frac{1}{N} {\sum_{i}} \mu_{r;i},
\end{equation}
where $\mu_{r;i}$ is the Ly$\alpha$ intensity in the images at pixel $i$ for the bin $r$.
$N$ is the number of all LAE-pixel pairs separated by $r$ within a certain bin.
The obtained $\xi_{\mathrm{IM}} (r)$ is not the dimensionless cross-correlation function, but the cross-correlation function with the unit of intensity.
Note that this 2D intensity mapping analysis method is basically the same as image stacking. 
We set six bins logarithmically spaced between $1\farcs5$ and $40''$.
In our calculation, image pixels can contribute more than once to the cross-correlation functions. 
However, the contribution of such pixels is unlikely to be significant, since the minimum distance between our LAEs is about $20''$ and most of them are separated by more than $100''$.
Statistical uncertainties on the measurements are computed by a bootstrap resampling method.
We randomly select LAEs with the number same as that of our LAEs, allowing a duplication.
We create 1,000 LAE samples by the random selection, and derive a 1$\sigma$ standard deviation that is referred to as a 1$\sigma$ error.

\subsection{Tests for systematic errors} \label{subsec:Non-LAE}
There are a number of systematic uncertainties that may produce spurious extended sources in the measurement of diffuse emissions.
For example, a largely-extended point-spread function (PSF) and flat-fielding / sky-subtraction systematics may affect the extended profile.
We need to carefully evaluate total uncertainties from these systematics.
To test the systematics in the cross-correlation between the Ly$\alpha$ intensity map and the LAEs, we derive a cross-correlation function between the Ly$\alpha$ intensity map and real foreground objects residing at neither $z=5.7$ nor $6.6$.
We refer to these objects as Non-LAEs.
The Non-LAEs are not correlate with Ly$\alpha$ emission at neither $z = 5.7$ nor $6.6$.

For a catalog of the Non-LAEs, we use the $g$-band dropout galaxy catalog taken from the ``Great Optically Luminous Dropout Research Using Subaru HSC'' (GOLDRUSH; \citealt{Ono2018}; \citealt{Harikane2018}) program.
These Non-LAEs ($g$-dropout galaxies) are selected mainly by probing the redshifted Lyman break feature and relatively blue UV continuum slope based on the broadband color selection criteria of $g-r > 1.0$, $r-i < 1.0$, and $g-r > 1.5(r-i)+0.8$. Based on their selection completeness simulations, the expected redshifts of the Non-LAEs are $z \simeq 3.8 \pm 0.5$, which is significantly lower than those of our LAEs, indicating that their Ly$\alpha$ emission is not captured in our NB images.  
Thanks to the wide redshift coverage, the total number of the Non-LAEs is about 10000 in each field.
Because the systematics like the largely-extended PSF should depend on the magnitude of the sources, 
we randomly select the Non-LAEs with the NB magnitude distribution same as that of our LAEs.
Since we should reduce the statistical errors of this analysis originated from the Non-LAE samples as much as possible, we construct a sample of the Non-LAEs whose number is 3--13 times larger than that of the LAEs in each field.
We measure the cross-correlation with Non-LAEs in the same manner as the one with LAEs, and evaluate the systematic uncertainties.

\begin{figure*}[htb!]
\plotone{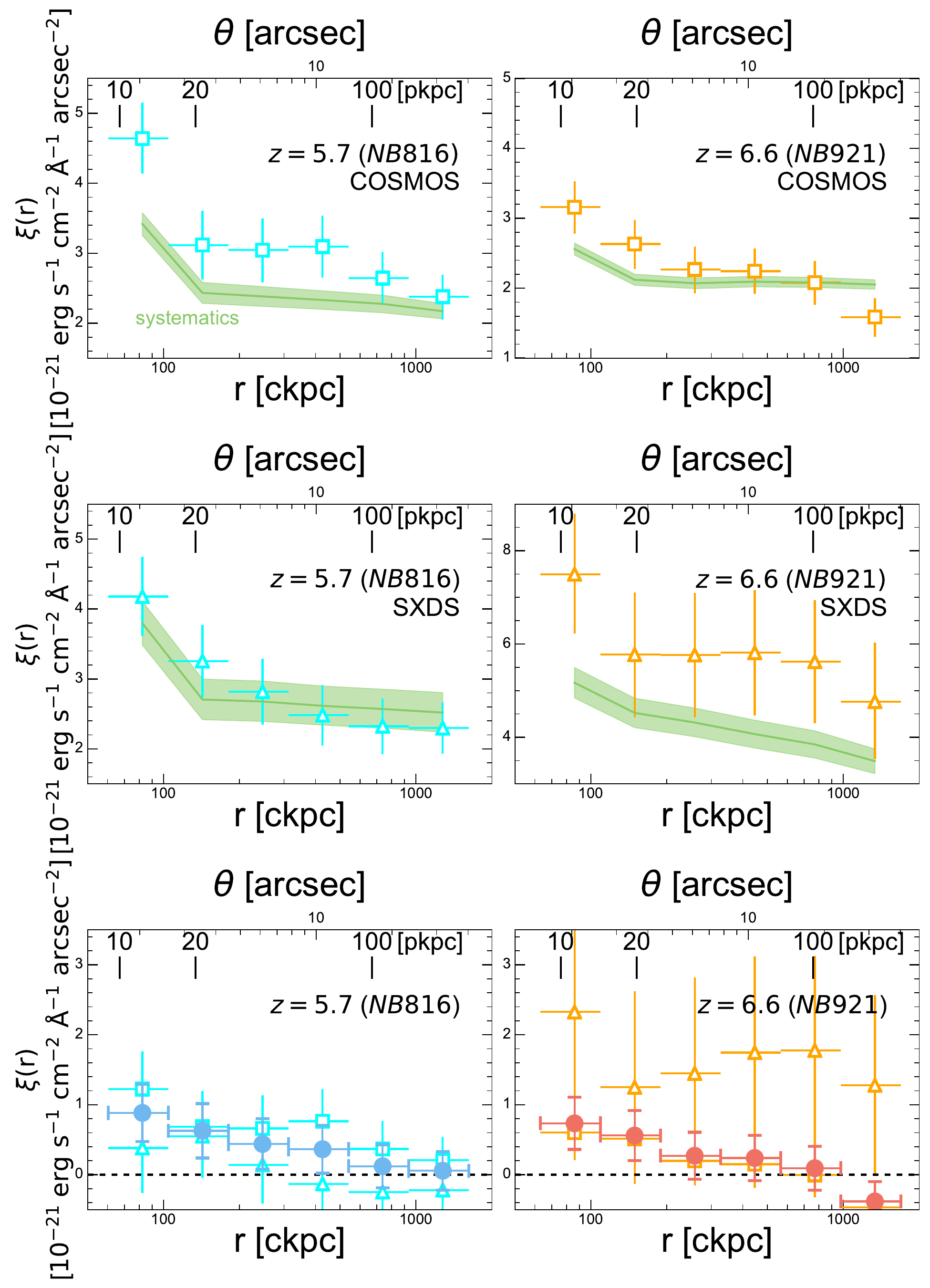}
\caption{
Cross-correlation function between the LAEs (Non-LAEs) and the NB image in each field.
The open squares in the top left and right panels show the present the cross-correlation function of the LAEs at $z = 5.7$ and $6.6$ in the UD-COSMOS field, respectively. 
The open triangles in the middle panels are the same as the top panels, but the results for the UD-SXDS field.
The green lines denote the cross-correlation functions between the Non-LAEs and the NB image that represent the amount of the total systematics (See text).
The green shade regions indicate the 1 $\sigma$ errors.
The bottom panels show the cross-correlation functions of the LAEs  after subtracting the systematics.
The open squares and triangles in the left (right) panel denote the results of $z=5.7$ ($6.6$) in the UD-COSMOS and UD-SXDS fields, respectively.
The filled circles in the left (right) panel represent the weighted mean values of the results of $z=5.7$ ($6.6$) in the UD-COSMOS and UD-SXDS fields that are calculated with the weights defined by the inverse values of the 1 $\sigma$ errors.
}
\label{fig:ALL_NB_result}
\end{figure*}

\subsection{Cross-correlation between LAEs and Ly$\alpha$ intensity}
\label{subsec:Lya}
In Figure \ref{fig:ALL_NB_result}, we show the cross-correlation between the LAEs (Non-LAEs) and the Ly$\alpha$ intensity map.
The cross-correlations with the Non-LAEs have positive values even at a large scale.
It is probably because the images in S18A data release have small residuals that are left in the sky-background subtraction, while the various techniques are applied to remove the residuals.
Comparing the cross-correlations of the LAEs and the Non-LAEs, we estimate the amount of the small residuals.
Figure \ref{fig:ALL_NB_result} indicates that the amplitude of the cross-correlation function of the LAEs is higher than those of Non-LAEs at each redshift in each field, although the $z=5.7$ LAEs in UD-SXDS show no clear excess especially at a large scale of $>200$ ckpc. 
Because the cross-correlation functions of the LAEs mostly exceed those of the Non-LAEs, we confirm that the systematic uncertainties alone do not explain the cross-correlation functions of the LAEs, but real signals of spatially-extended Ly$\alpha$ emission.

To evaluate the spatially-extended Ly$\alpha$ emission quantitatively, we subtract the cross-correlation function of the Non-LAEs from the one of the LAEs 
as shown in the bottom panels of Figure \ref{fig:ALL_NB_result}. 
The calculated values are presented in Table \ref{table:Lya_cross_correlation}.
We obtain weighted mean values of the results in the UD-COSMOS and UD-SXDS fields that are calculated with the weights defined by the inverse values of the 1$\sigma$ errors. 
The uncertainties of the measurements are estimated on the basis of the error propagation.
Our results show spatially extended emission at the scale of $\gtrsim 100$ ckpc.

To examine a possibility that this spatially extended emission in the weighted mean of the cross-correlation functions is caused by other systematics that are not corrected in the analysis above, we also investigate the cross-correlation between the LAEs and the intensity map taken with another filter. 
Here we use $g$-band images taken from the HSC-SSP survey S18A data release (Section \ref{subsec:Images}).
The 5$\sigma$ limiting magnitude of the $g$-band images is $\sim 27$ mag.
Because the $g$-band filter covers a wavelength range shorter than the observed-frame wavelength of Ly$\alpha$ emission at $z=5.7$ and $6.6$, the LAEs in our samples should not be detected in the $g$-band images.
We thus expect that there are no signals in the cross-correlation between the LAEs and the $g$-band images.
One can test a reliability of our results with the systematics subtraction, investigating departures from zero in the cross-correlation between the LAEs and the $g$-band images. 
Performing this test, we derive the cross-correlation function between the LAEs and the $g$-band images in the UD-COSMOS and UD-SXDS fields, respectively, that are corrected for the systematics in the same manner as our results. 
In this test, the Non-LAE sample consists of $g$-dropouts whose $g$-band fluxes are mostly negligibly small.
We confirm that the cross-correlation function is consistent with zero within the errors, and that there is no significant systematics mimicking the spatially-extended Ly$\alpha$ emission in our results of Figure 
\ref{fig:ALL_NB_result}.

Note that the cross-correlations of the $z=5.7$ ($z=6.6$) LAEs after subtracting the systematics in COSMOS (SXDS) appear to be larger than that in the other field with large uncertainties. 
Although we carefully control the systematics by using the cross-correlations of the Non-LAEs, this difference might be induced by some unknown uncertainty that is not corrected in our careful analysis. 
Conceivably this difference might be caused by an environmental effect, although in that case the dependency of the cross-correlation on the galaxy overdensity appears to be opposite to that reported in previous work \citep{Matsuda2012}. 
A more detailed examination on this issue is beyond the scope of this paper; we would like to derive cross-correlations of LAEs that are detected with other NBs whose central wavelengths are shorter than ours, to check whether similar differences are seen or not, which may be helpful to address this issue, in a forthcoming paper.

\begin{figure*}[htb!]
\plotone{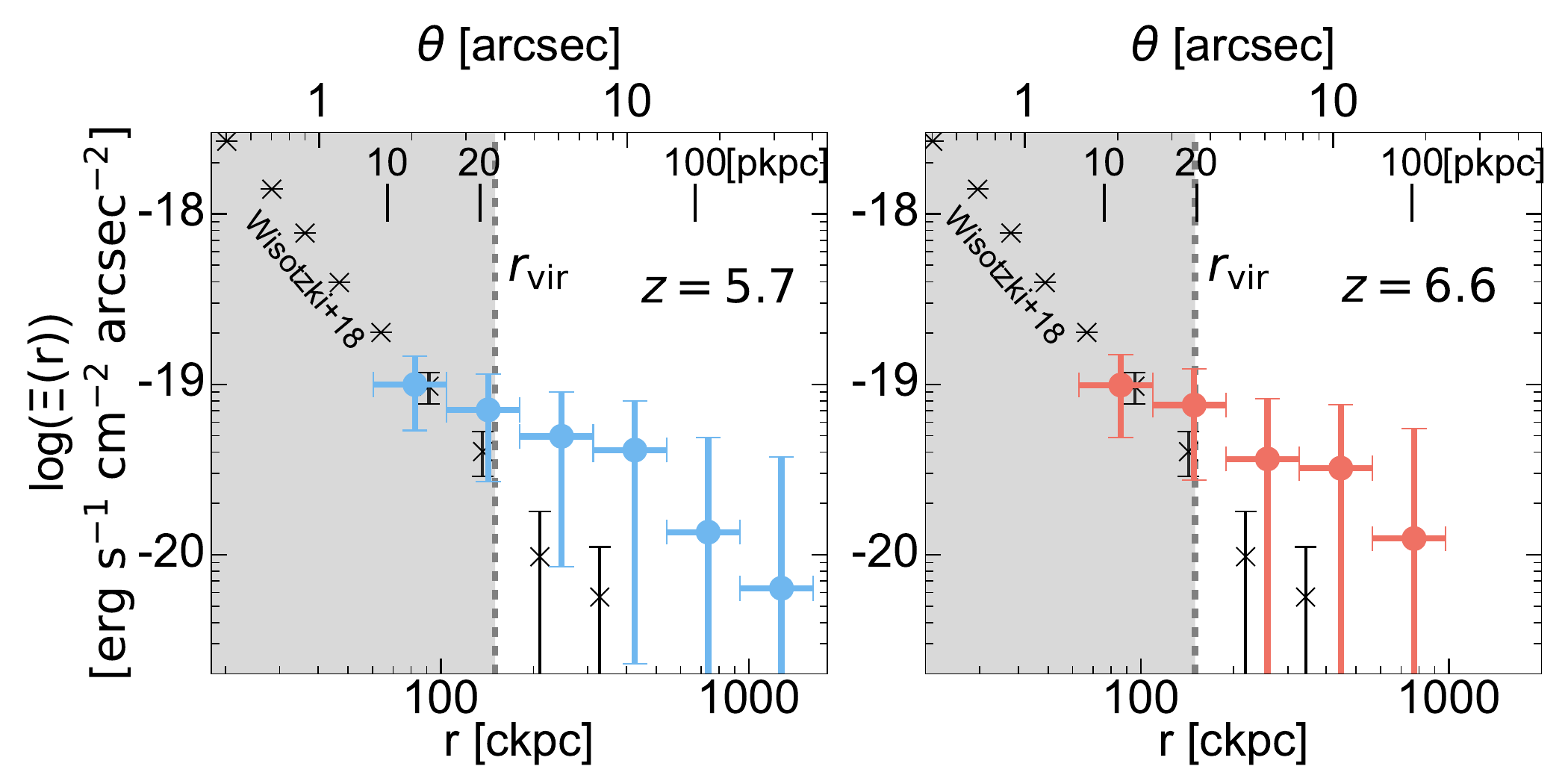}
\caption{
Cross-correlation function between the LAEs and the Ly$\alpha$ emission intensity after subtracting the systematics.
The filled circles in the left (right) panel represent the weighted mean of the results in UD-COSMOS and UD-SXDS fields of $z= 5.7$ ($6.6$). 
The S/Ns of the outer three data points are lower than $1$ as presented in Table \ref{table:Lya_cross_correlation}.
The black crosses correspond to the stacked Ly$\alpha$ radial profile at $z=5$--$6$ that is taken from the observational results of MUSE \citep{Wisotzki2018}.
The gray dashed lines show the DMH virial radius ($r_\mathrm{vir}$) of the LAEs in our samples whose DMH masses are estimated to be $M_\mathrm{h} \sim 10^{11} M_\odot$.
}
\label{fig:NB_MUSE_log}
\end{figure*}

In Figure \ref{fig:NB_MUSE_log}, we show the weighted mean values of the cross-correlation functions at $z = 5.7$ and $6.6$. 
Here, we convert $\xi$(r) [erg s$^{-1}$ cm$^{-2}$ \AA$^{-1}$ arcsec$^{-2}$] to the Ly$\alpha$ flux cross-correlation function $\Xi$(r) [erg s$^{-1}$ cm$^{-2}$ arcsec$^{-2}$] by multiplying the FWHM of NB filter (FWHM$_{\mathrm{NB}}$).
\begin{equation}
\Xi(r) = \xi_{\mathrm{IM}}(r) \times \mathrm{FWHM_{NB}}
\label{eq:Xi_r}
\end{equation}
The obtained results are presented in Table \ref{table:Lya_cross_correlation}. 
For comparison, in Figure \ref{fig:NB_MUSE_log}, we also present the Ly$\alpha$ radial profile of relatively luminous LAEs at similar redshifts of $z=5$--$6$ with Ly$\alpha$ luminosities of $> 10^{42}$ erg s$^{-1}$ obtained by MUSE observations (Extended Data Figure 5 in \citealt{Wisotzki2018}).
As shown in Figure \ref{fig:NB_MUSE_log}, the MUSE observations identify Ly$\alpha$ emission at the small scale of $< 150$ ckpc beyond the errors. 
In this scale, the Ly$\alpha$ intensity radial profiles of our results are consistent with those of the MUSE observations both at $z=5.7$ and $6.6$.
We also check previous MUSE results for Ly$\alpha$ radial profiles around individual high-$z$ LAEs reported by \cite{Leclercq2017a}. In their sample, the \#547 LAE has a comparable Ly$\alpha$ luminosity ($10^{42.77}$ erg s$^{-1}$) and is located at a similar redshift ($z=5.98$) to our samples. We find that our results are also consistent with the Ly$\alpha$ radial profile of this individual LAE at the small scale of $< 150$ ckpc.

At the larger scale of $\gtrsim$ 100 ckpc, our results indicate the possible existence of very faint spatially-extended Ly$\alpha$ emission more largely beyond the errors, 
which is not probed with the MUSE results. 
We estimate the detection confidence levels of the spatially-extended Ly$\alpha$ emission at the scale of $\gtrsim$ 100 ckpc ($\simeq$ 80--400 ckpc) to be the 2.6 $\sigma$ and 2.2$\sigma$ levels at $z=5.7$ and $6.6$, respectively, based on Fisher's method.\footnote{The detection confidence levels of the Ly$\alpha$ cross-correlations at the scale of $\gtrsim$ 100 ckpc are estimated to be 
2.9$\sigma$ for the $z=5.7$ COSMOS sample, 
0.4$\sigma$ for the $z=5.7$ SXDS sample, 
1.6$\sigma$ for the $z=6.6$ COSMOS sample, 
and 
2.1$\sigma$ for the $z=6.6$ SXDS sample. 
} 
Figure \ref{fig:NB_MUSE_log} also indicates a hint of the extended Ly$\alpha$ emission up to $\sim 1$,$000$ ckpc.
Based on the halo occupation distribution models, LAEs in our sample are hosted by the DMHs with the average mass of $\log(\langle M_\mathrm{h} \rangle / M_\odot) = 10.8^{+0.3}_{-0.5}$ ($11.1^{+0.2}_{-0.4}$) at $z = 5.7$ ($6.6$) with a Ly$\alpha$ duty cycle of 1\% or less \citep{Ouchi2018}.
The DMH virial radius a galaxy whose halo mass is $\log(M_\mathrm{h}/M_\odot) = 11$ is $r_\mathrm{vir} \sim 150$ ckpc \citep{Mo2002}. 
By the comparison with the DMH virial radius of $r_\mathrm{vir} \sim 150$ ckpc, our study has shown the marginal detection of the spatially-extended Ly$\alpha$ emission around and beyond the DMH virial radius at the $\simeq 3\sigma$ ($\simeq 2\sigma$) level at $z=5.7$ ($6.6$).\footnote{If we calculate the detection confidence levels of the spatially extended Ly$\alpha$ emission only beyond the virial radius, we obtain $1.9\sigma$ at $z=5.7$ and $1.5 \sigma$ at $z=6.6$.}

\begin{deluxetable*}{ccccc} 
\tablecolumns{5} 
\tablewidth{0pt} 
\tablecaption{
Ly$\alpha$ flux cross-correlation functions 
for the $z=5.7$ and $z=6.6$ LAEs \label{table:Lya_cross_correlation}}
\tablehead{
    \colhead{$\theta$} 
    & \colhead{$\xi(r)_\mathrm{COSMOS}$} 
    & \colhead{$\xi(r)_\mathrm{SXDS}$} 
    & \colhead{$\xi(r)$} 
    & \colhead{$\Xi(r)$} 
    \\ 
    \colhead{(arcsec)} 
    & \colhead{($10^{-21}$ erg s$^{-1}$ cm$^{-2}$ {\AA}$^{-1}$ arcsec$^{-2}$)} 
    & \colhead{($10^{-21}$ erg s$^{-1}$ cm$^{-2}$ {\AA}$^{-1}$ arcsec$^{-2}$)}  
    & \colhead{($10^{-21}$ erg s$^{-1}$ cm$^{-2}$ {\AA}$^{-1}$ arcsec$^{-2}$)}  
    & \colhead{($10^{-20}$ erg s$^{-1}$ cm$^{-2}$ arcsec$^{-2}$)} 
    \\ 
    \colhead{(1)} & 
    \colhead{(2)} & 
    \colhead{(3)} & 
    \colhead{(4)} & 
    \colhead{(5)}
}
\startdata 
    \multicolumn{5}{c}{$z = 5.7$}\\ \hline
     2.05 & $1.225^{+0.535}_{-0.529}$ & $0.382^{+0.648}_{-0.643}$ & 0.883 $\pm$ 0.409 & 9.98 $\pm$ 4.62 \\ 
     3.54 & $0.685^{+0.509}_{-0.515}$ & $0.550^{+0.591}_{-0.593}$ & 0.627 $\pm$ 0.389 & 7.09 $\pm$ 4.39 \\ 
     6.13 & $0.662^{+0.474}_{-0.483}$ & $0.140^{+0.556}_{-0.551}$ & 0.438 $\pm$ 0.363 & 4.95 $\pm$ 4.10 \\ 
     10.6 & $0.764^{+0.461}_{-0.462}$ & $-0.133^{+0.515}_{-0.515}$ & 0.364 $\pm$ 0.344 & 4.11 $\pm$ 3.88 \\ 
     18.3 & $0.368^{+0.396}_{-0.403}$ & $-0.247^{+0.489}_{-0.483}$ & 0.120 $\pm$ 0.309 & 1.36 $\pm$ 3.50 \\ 
     31.7 & $0.207^{+0.335}_{-0.346}$ & $-0.221^{+0.465}_{-0.459}$ & 0.056 $\pm$ 0.276 & 0.64 $\pm$ 3.12 \\ 
     \hline
     \multicolumn{5}{c}{$z = 6.6$}\\ \hline
     2.05 & $0.600^{+0.378}_{-0.389}$ & $2.326^{+1.342}_{-1.309}$ & 0.734 $\pm$ 0.373 & 9.90 $\pm$ 5.04 \\ 
     3.54 & $0.512^{+0.356}_{-0.369}$ & $1.249^{+1.371}_{-1.382}$ & 0.559 $\pm$ 0.357 & 7.55 $\pm$ 4.81 \\ 
     6.13 & $0.197^{+0.338}_{-0.348}$ & $1.448^{+1.369}_{-1.374}$ & 0.270 $\pm$ 0.337 & 3.65 $\pm$ 4.55 \\ 
     10.6 & $0.150^{+0.334}_{-0.335}$ & $1.746^{+1.375}_{-1.379}$ & 0.239 $\pm$ 0.326 & 3.22 $\pm$ 4.40 \\ 
     18.3 & $-0.003^{+0.320}_{-0.323}$ & $1.776^{+1.348}_{-1.351}$ & 0.092 $\pm$ 0.314 & 1.24 $\pm$ 4.24 \\ 
\enddata
\tablecomments{
(1) Projected distances from the LAEs. 
(2) Cross-correlation between the LAEs and the Ly$\alpha$ intensity for the COSMOS field.
(3) Same as (2) but for the SXDS field.
(4) Same as (2) but for the average values of the two fields.
(5) Ly$\alpha$ flux cross-correlation functions obtained from Eq. (\ref{eq:Xi_r}). 
}
\end{deluxetable*}

\section{Discussion} \label{sec:discussion}
\subsection{Extended Ly$\alpha$ emission}
\label{sec:extended_lya_emission}
\begin{figure}[h!]
\plotone{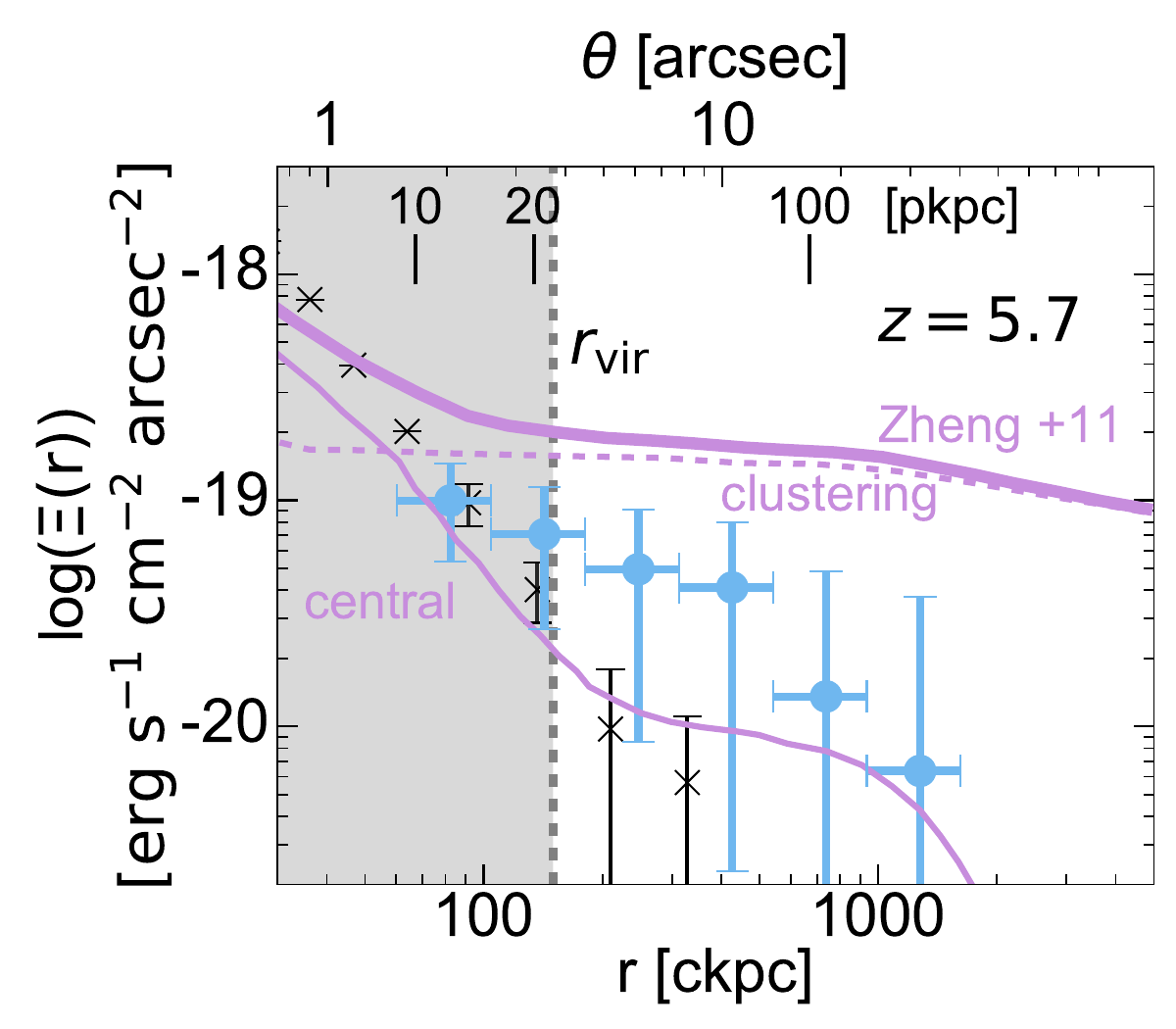}
\caption{
Ly$\alpha$ emission radial profiles of the observational results that are compared with those are predicted with the radiation-hydrodynamic reionization simulation \citep{Zheng2011a}.
The thick solid line denotes a radial profile of the stacked $z=5.7$ LAE images in the simulation (See the text for details about the simulation).
The simulated Ly$\alpha$ radial profile is produced by two sources of Ly$\alpha$ photons, the central LAE (thin solid line) and the clustered sources around the central LAE (dashed line).
The blue circles and the black are same as those in the left panel of 
Figure \ref{fig:NB_MUSE_log}. 
}
\label{fig:Zheng}
\end{figure}

By the cross-correlation analysis, we tentatively identify the Ly$\alpha$ emission spatially extending beyond the scale of $r_\mathrm{vir}$.
Theoretical studies predict that {\sc Hi} gas in the CGM and IGM resonantly scatters Ly$\alpha$ photons that escape from the inter-stellar medium (ISM) of star-forming galaxies.
\citet{Zheng2010} perform a radiation-hydrodynamic reionization simulation in a cosmological volume, and present a physical model of Ly$\alpha$ emission observed around LAEs.
The radiative transfer simulation is conducted to explain the observed properties of LAEs at $z\sim 5.7$ in the Subaru/XMM-Newton Deep Survey (SXDS; \citealt{Ouchi2008}).
This simulation includes physical processes of Ly$\alpha$ photons scattered by the CGM and IGM.
On the basis of the simulation, \citet{Zheng2011a} present Ly$\alpha$ radial profiles of stacked images of model LAEs in the simulation. The model LAEs are residing in halos with masses of $M_\mathrm{h} \sim 10^{11} M_\odot$.
Figure \ref{fig:Zheng} compares Ly$\alpha$ radial profiles of the simulation results and our observational results for the LAEs at $z=5.7$. 
For comparison, we use the simulation results of {\lqq}the bright half of the sources{\rqq} presented in Figure 6 of \citet{Zheng2011a}, because these sources have Ly$\alpha$ luminosities comparable to those of our LAEs identified in the HSC observations.

\citet{Zheng2011a} present that the Ly$\alpha$ radial profile of the simulation consists of two components whose Ly$\alpha$ sources are star-forming regions of the LAE (called {\lqq}central LAE{\rqq}) and clustered sources around the LAE (called {\lqq}clustering{\rqq}), i.e. neighboring galaxies including satellites. 
Figure \ref{fig:Zheng} also shows the total Ly$\alpha$ radial profile, a sum of the two-component Ly$\alpha$ radial profiles. 
Figure \ref{fig:Zheng} indicates that our observational results agree neither with the total profile nor the clustered-source profile.
This is probably because we mask out the pixels in which LAEs are detected at S/N$>5$ (Section \ref{subsec:Images}). 
In other words, in our measurements, the contribution from satellite LAEs down to our Ly$\alpha$ luminosity limit of about $10^{43}$ erg s$^{-1}$ is subtracted. 
Note that, based on the previous results of very deep LAE searches, LAEs whose Ly$\alpha$ luminosities are well below our detection limits do exist \citep{Drake2017}, and such faint LAEs would contribute to some extent to our obtained Ly$\alpha$ cross-correlation function.
However, Figure \ref{fig:Zheng} suggests that the Ly$\alpha$ radial profile shapes are different between the clustered source {\lqq}clustering{\rqq} and our observational results. 
In Figure \ref{fig:Zheng}, the {\lqq}clustering{\rqq} profile decreases only by a factor less than 2 at 150--1000 ckpc, while our observational result profile changes by an order of magnitude beyond the errors. 
This profile shape difference suggests that such faint LAEs are unlikely to be dominant contributors to the Ly$\alpha$ cross-correlation. 
In Figure \ref{fig:Zheng}, the Ly$\alpha$ radial profile of our observational results with the error bars prefers the model of {\lqq}central LAE{\rqq} in the simulation.
However, the Ly$\alpha$ radial profile of our observational results is higher than the one of the {\lqq}central LAE{\rqq} component, albeit with the large observational uncertainties.
The Ly$\alpha$ radial profile of the observations exceeding the {\lqq}central LAE{\rqq} component may be produced by other physical processes that are not included in the simulation of \citet{Zheng2010}, such as a cold-gas stream and galactic outflow that generate Ly$\alpha$ photons by collisional excitation processes.

To understand physical origins of the spatially-extended Ly$\alpha$ emission at the large scale, numerical simulations with various possible physical processes are needed \citep{Sadoun2019}.
We also need Ly$\alpha$ intensity mapping measurements with small uncertainties to distinguish the various possible physical models.
More Ly$\alpha$ intensity mapping studies should be conducted with the existing instruments including Subaru/HSC. Moreover, we expect to obtain conclusive Ly$\alpha$ intensity mapping results by the on-going and forthcoming projects such as Hobby-Eberly Telescope Dark Energy Experiment (HETDEX; \citealt{Hill2008}), Wide-Field InfrarRed Survey Telescope (WFIRST; \citealt{Spergel2015a}), and Spectro-Photometer for the History of the Universe Epoch of Reionization and Ices Explorer (SPHEREx; \citealt{Dore2014}).

\subsection{Cosmic Reionization}
\begin{figure}[h!]
\plotone{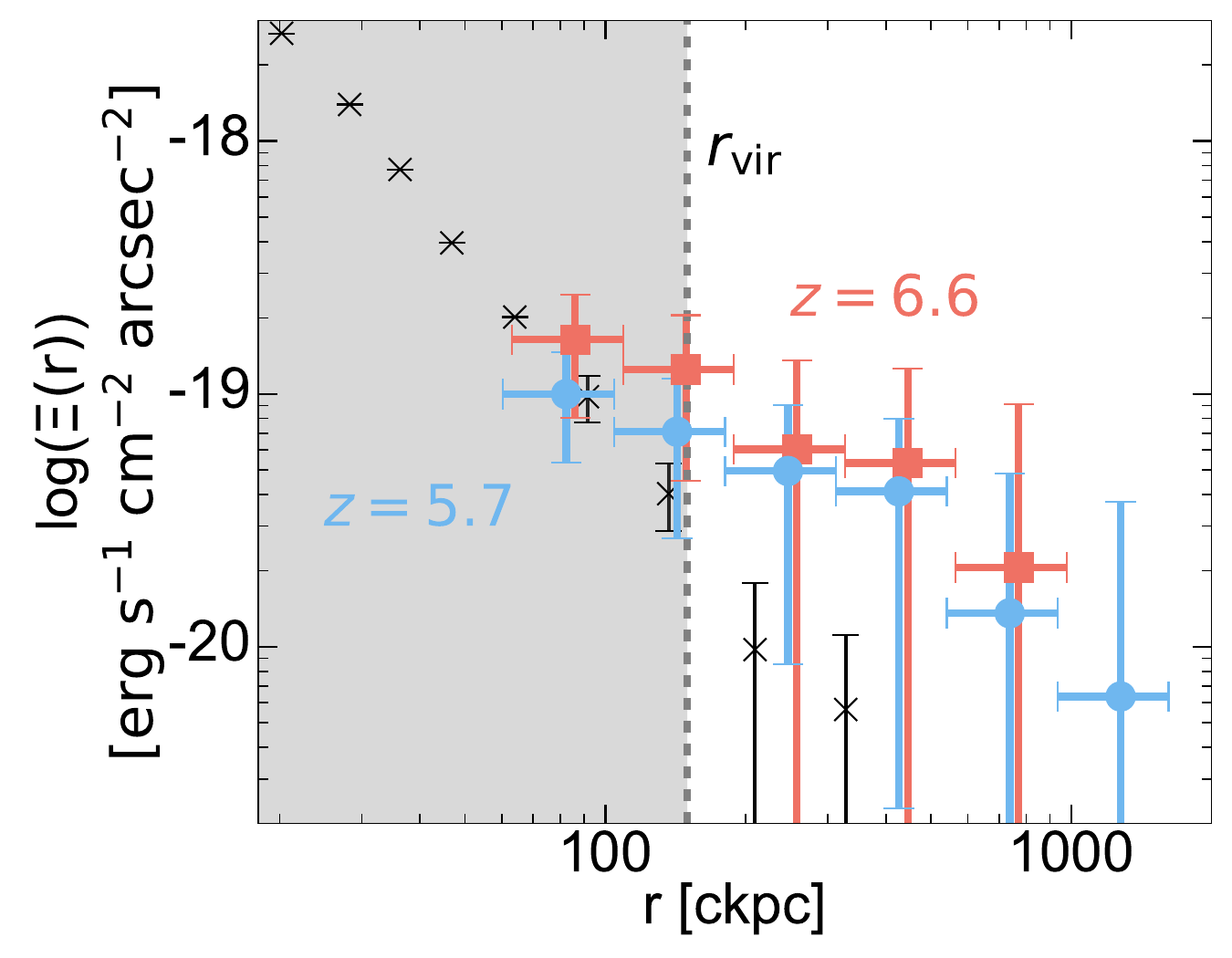}
\caption{
Comparison between the cross-correlation functions of the LAEs at $z=5.7$ and $6.6$.
The blue circles and red squares represent the Ly$\alpha$ cross-correlation functions at $z=5.7$ and $6.6$, respectively.
We match the Ly$\alpha$ radial profile at $z=6.6$ to the one at $z=5.7$, correcting for the cosmological surface brightness dimming effect.
}
\label{fig:Reionization}
\end{figure}

Many observational studies have claimed the increase of the IGM neutral hydrogen fraction $x_{\mathrm{\sc H_I}}$ with the decrease of the Ly$\alpha$ luminosity function from $z\sim 6$ to $7$ and beyond that can be explained by the increase of the Ly$\alpha$ damping wing absorption given by the neutral IGM (e.g. \citealp{Ouchi2010, Goto2011, Kashikawa2011a}). \citet{Jeeson-Daniel2012} predict that the Ly$\alpha$ radial profile of LAEs flattens towards the early epoch of cosmic reionization with a high $x_\mathrm{\sc H_I}$, due to the IGM neutral hydrogen scattering Ly$\alpha$ photons. Because we have the observational data of the Ly$\alpha$ radial profiles at 
the epoch of reionization ($z=6.6$) and the post reionization epoch ($z=5.7$), we examine whether the flattening is found in our observational data.
It should be noted that the Ly$\alpha$ luminosity ranges of our LAEs at $z=5.7$ and $6.6$ are comparable, $\log ( L_{\mathrm{Ly\alpha}}/{\rm [erg\ s}^{-1}{\rm]} ) = 42.0-43.8$ and $42.3-44.0$, respectively \citep{Konno2018}.
The difference between the median values of the Ly$\alpha$ luminosity of the LAEs at $z=5.7$ and $6.6$ is small (within a factor of $3$).
Figure \ref{fig:Reionization} compares our observational results at $z=5.7$ and $6.6$. Here we apply corrections for the surface brightness dimming effect in our $z=6.6$ results, and carry out the comparison at $z=5.7$.
Figure \ref{fig:Reionization} indicates that our observational results at $z=5.7$ and $6.6$ are similar, and that there is no signature of the flattening of the Ly$\alpha$ radial profile towards high-$z$ beyond the errors. Because the error bars, especially at a large scale, of our observational results are probably too large to identify the Ly$\alpha$ radial profile flattening caused by the cosmic reionization, one should investigate a Ly$\alpha$ radial profile slope with the large observational data that will be taken by the on-going and forthcoming programs such as HETDEX, WFIRST, and SPHEREx discussed in Section \ref{sec:extended_lya_emission}.

\section{Summary} \label{sec:summary}
In this paper, we report the first evidence of the Ly$\alpha$ emission around the star-forming galaxies largely extended around and probably even beyond the DMH virial radius ($r_\mathrm{vir} \sim 150$ ckpc  for $M_\mathrm{h} \sim 10^{11} M_\odot$) at $z=5.7$ and $6.6$ based on the Subaru/HSC intensity mapping analysis of the cross correlation between the Ly$\alpha$ intensity map and the positions of LAEs.
The major results of this paper are summarized below.

1) Using the largest $z \gtrsim 5$ LAE catalogs to date and the large-area ($\sim 4$ deg$^2$) NB imaging data, we conduct the cross-correlation intensity mapping analysis with the LAEs and the Ly$\alpha$ intensity map.
We conduct extensive analyses evaluating systematics of large-scale PSF wings, sky subtractions, and unknown errors.
Confirming that our Ly$\alpha$ cross-correlations corrected for the systematics produce no spurious detections by the careful tests, we have reported a possible detection of very diffuse Ly$\alpha$ emission at the $\simeq 3\sigma$ ($\simeq 2\sigma$) significance level at the distance of $\gtrsim$ 100 ckpc from the LAEs at $z = 5.7$ ($6.6$), around and beyond a virial radius of star-forming galaxies with $M_\mathrm{h} \sim 10^{11} M_\odot$. The diffuse Ly$\alpha$ emission possibly extends up to $1$,$000$ ckpc with the surface brightness of $10^{-20}$--$10^{-19}$ erg s$^{-1}$ cm$^{-2}$ arcsec$^{-2}$.
In this analysis, we confirm that the small-scale ($<150$ ckpc) Ly$\alpha$ radial profiles of LAEs in our Ly$\alpha$ intensity maps are consistent with those obtained by recent MUSE observations (\citealt{Leclercq2017a}; \citealt{Wisotzki2018}).

2) Comparisons with numerical simulations suggest that the large-scale ($\sim 150$--$1$,$000$ ckpc) Ly$\alpha$ emission are not explained by unresolved faint sources of neighboring galaxies including satellites, but by a combination of Ly$\alpha$ photons emitted from the central LAE and other unknown sources, such as a cold-gas stream and galactic outflow.

3) Although theoretical studies predict that the Ly$\alpha$ radial profile is flattened towards the early epoch of cosmic reionization, we find no evolution in the Ly$\alpha$ radial profiles of our LAEs from $z=5.7$ to $6.6$. However, the moderately large errors may not allow us to identify the flattening of the Ly$\alpha$ radial profiles towards high redshift.

The evolution of the Ly$\alpha$ radial profiles at the EoR should be investigated by very large-area surveys with the on-going and forthcoming programs such with Subaru/HSC, HETDEX, WFIRST, and SPHEREx.

\section*{Acknowledgments} \label{sec:acknow}
The Hyper Suprime-Cam (HSC) collaboration includes the astronomical communities of Japan and Taiwan, and Princeton University. 
The HSC instrumentation and software were developed by the National Astronomical Observatory of Japan (NAOJ), the Kavli Institute for the Physics and Mathematics of the Universe (Kavli IPMU), the University of Tokyo, the High Energy Accelerator Research Organization (KEK), the Academia Sinica Institute for Astronomy and Astrophysics in Taiwan (ASIAA), and Princeton University. 
Funding was contributed by the FIRST program from Japanese Cabinet Office, the Ministry of Education, Culture, Sports, Science and Tech- nology (MEXT), the Japan Society for the Promotion of Science (JSPS), Japan Science and Technology Agency (JST), the Toray Science Foundation, NAOJ, Kavli IPMU, KEK, ASIAA, and Princeton University. 
The \textit{NB816} filter was supported by Ehime University (PI: Y. Taniguchi). The \textit{NB921} filter was supported by KAKENHI (23244025) Grant-in-Aid for Scientific Research (A) through the Japan Society for the Promotion of Science (PI: M. Ouchi). 
This paper makes use of software developed for the Large Synoptic Survey Telescope. We thank the LSST Project for making their code available as free software at http://dm.lsst.org. 
The Pan-STARRS1 Surveys (PS1) have been made possible through contributions of the Institute for Astronomy, the University of Hawaii, the Pan-STARRS Project Office, the Max-Planck Society and its participating institutes, the Max Planck Institute for Astronomy, Heidelberg and the Max Planck Institute for Extraterrestrial Physics, Garching, The Johns Hopkins University, Durham University, the University of Edinburgh, Queen’s University Belfast, the Harvard-Smithsonian Center for Astrophysics, the Las Cumbres Observatory Global Telescope Network Incorporated, the National Central University of Taiwan, the Space Telescope Science Institute, the National Aeronautics and Space Administration under Grant No. NNX08AR22G issued through the Planetary Science Division of the NASA Science Mission Directorate, the National Science Foundation under Grant No. AST-1238877, the University of Maryland, and Eotvos Lorand University (ELTE). 
This work is supported by World Premier International Research Center Initiative (WPI Initiative), MEXT, Japan, and KAKENHI (15H02064, 17KK0098, 17H01110, 17H01114, 17H04831, and 20H00180) Grant-in-Aid for Scientific Research (A) through Japan Society for the Promotion of Science.
Based on data collected at the Subaru Telescope and retrieved from the HSC data archive system, which is operated by Subaru Telescope and Astronomy Data Center at National Astronomical Observatory of Japan.

\bibliographystyle{apj}
\bibliography{Kakuma_2019}

\end{document}